\documentclass[letterpaper,english,reprint, aps]{revtex4-1}
\usepackage[T1]{fontenc}
\usepackage[utf8]{inputenc}
\usepackage{amsmath}
\usepackage{amssymb}
\usepackage{graphicx}

\makeatletter

\pdfpageheight\paperheight
\pdfpagewidth\paperwidth

\usepackage{amsmath}

\newcommand{\Mod}[1]{\ (\mathrm{mod}\ #1)}

\makeatother

\usepackage{babel}
\begin{document}
\preprint{APS/123-QED}
\title{Arbitrary optical wave evolution with Fourier transforms and phase
masks}
\author{Víctor J. López-Pastor}
\affiliation{Max Planck Institute for the Science of Light, Staudtstr. 2, 91058
Erlangen }
\author{Jeff S. Lundeen}
\affiliation{Department of Physics and Centre for Research in Photonics, University
of Ottawa }
\author{Florian Marquardt}
\affiliation{Max Planck Institute for the Science of Light, Staudtstr. 2, 91058
Erlangen}
\date{\today}
\begin{abstract}
A large number of applications in classical and quantum photonics
require the capability of implementing arbitrary linear unitary transformations
on a set of optical modes. In a seminal work by Reck\emph{ et al.
}\citep{reck_experimental_1994} it was shown how to build such multiport
universal interferometers with a mesh of beam splitters and phase
shifters, and this design became the basis for most experimental implementations
in the last decades. However, the design of Reck\emph{ et al. }is
difficult to scale up to a large number of modes, which would be required
for many applications. Here we present a constructive proof that it
is possible to realize a multiport universal interferometer on $N$
modes with a succession of $6N$ Fourier transforms and $6N+1$ phase
masks, for any even integer $N$. Furthermore, we provide an algorithm
to find the correct succesion of Fourier transforms and phase masks
to realize a given arbitrary unitary transformation. Since Fourier
transforms and phase masks are routinely implemented in several optical
setups and they do not suffer from the scalability issues associated
with building extensive meshes of beam splitters, we believe that
our design can be useful for many applications in photonics.
\end{abstract}
\maketitle
\emph{Introduction.—} The ability to arbitrarily transform an optical
mode has applications spanning communications, imaging, and information
processing. Such a transformation that is lossless and linear is described
by a unitary matrix $U$, mapping a basis of $N$ input modes onto
a basis of $N$ output modes. Since any such matrix has $N^{2}$ free
parameters, a method for its implementation must have at least $N^{2}$
controllable parameters, which is an experimentally challenging scaling.
One implementation method is based on optical Fourier transforms (FT)
\citep{morizur_programmable_2010,armstrong_programmable_2012,labroille_efcient_2014}.
In this paper, we show that only $N^{2}$ controllable parameters
are needed to implement an arbitrary unitary transformation on $N$
modes using FTs. What is more, we introduce a deterministic algorithm
to design an arbitrary unitary transformation based on this method.

General variable control of modal unitary transformations will have
applications across optics. For example, in fiber optic communications,
spatial multiplexing will require transforming between an array of
Gaussian profile modes from, say, a ribbon of single-mode fibers to
the non-Gaussian spatial modes of one multimode fiber \citep{Bozinovic20131545}.
Routing of optical channels requires a reconfigurable network, described
by a unitary \citep{zhuang_programmable_2015,perez_multipurpose_2017}.
Information processing with optical networks takes advantage of the
ultra-low latency and ultra-high clock speed of photonic waveguides.
Capitalizing on this allows for one to, for instance, concatenate
two unitary transformations to quickly multiply two matrices, a key
ingredient in a neural network \citep{shen_deep_2017,Steinbrecher2019}.
Another area of application, imaging, is, at its heart, a unitary
spatial transformation. General unitary transformations would enable
novel imaging functionalities, such as cancelling the optical scattering
that inhibits imaging through human tissue \citep{popoff2010measuring,popoff2010image}.
Image processing, such as noise reduction, sharpening, or compression,
could be done on the field itself, rather than the intensity recorded
by the sensor \citep{Silva160}. Turning to the area of quantum information,
a generalization of the $N=2$ qubit is a photon in a superposition
of $N$ modes, a qudit \citep{schaeff_experimental_2015,Larocque2017}.
In quantum cryptography, a protocol that uses qudits (and unitaries
on them) rather than qubits improves the robustness to noise \citep{bouchard2018experimental}.
Quantum computing logic gates, such as the controlled-NOT gate, can
be implemented using unitary transformations on photonic waveguide
modes \citep{Politi646}. Moreover, random walks in waveguide-network
transformations simulate a variety of quantum systems \citep{harris_quantum_2017},
such as molecules \citep{Peruzzo1500}. The problem of sampling the
output probability distribution when multiple photons traverse such
networks is hard to simulate in a classical computer and hence it
may be a viable path to achieve quantum supremacy with photonic devices,
as proved in \citep{aaronson_computational_nodate}. The underlying
reason is that sampling the bosonic statistics of the output photons
is linked to the problemn of estimating the permanent of a large unitary
matrix \citep{Broome794,Crespi2013,Spring798,Tillmann2013}, which
is $\textsc{\#P}$-hard \citep{VALIANT1979189}. These applications
motivate why the field is spending considerable effort to develop
controllable unitary transformations.

We now briefly outline these efforts and methods. While the first
methods to create arbitrary transformations were developed during
early radio and microwave engineering, Reck et al. introduced them
to optics in a seminal paper in 1994 \citep{reck_experimental_1994}.
One of the simplest tools available in optics is a phase shifter.
However, by themselves modal phase-shifters are insufficient to build
a general unitary transformation. In addition, one must use mode-mixing
elements such as beamsplitters. Reck et al. gave a prescription to
implement any chosen unitary on an array of beam modes by using a
triangle-shaped lattice of variable-reflectivity beamsplitters interleaved
with phase shifters. However, the complexity of this method meant
it was not demonstrated until an integrated optical implementation
over twenty years later \citep{Carolan711}. Soon after, a more compact
square lattice of beamsplitters and phase shifters was proposed and
implemented by Clements et al. \citep{clements_optimal_2016}. Since
the first implementation, a range of integrated optical platforms
have hosted proof-of-principle applications of these lattices \citep{crespi_suppression_2016,mennea_modular_2018,ribeiro_demonstration_2016}.
However, the fabrication and control complexity associated with this
method has, so far, limited demonstrations of arbitrary unitaries
to $N=6$ waveguides \citep{Carolan711}.

Before these integrated optical implementations, a different type
of mixer was proposed and implemented, a lens or curved mirror. The
latter elements enact an approximate Fourier transform of the spatial
field-distribution \citep{goodman2005introduction}. Using these,
a unitary is decomposed into a series of FTs interleaved with phase-shifters
\citep{morizur_programmable_2010,armstrong_programmable_2012,labroille_efcient_2014},
as shown in Fig. 1. The phase shifters are varied to implement a given
unitary, whereas the FTs do not change. An SLM of $N$ pixels per
side will inherently contain the $N^{2}$ control parameters required
to implement a unitary for a spatial mode that varies along a row
of pixels. Moreover, since they are based on television technology
(e.g., 4K resolution) they currently have up to $N^{2}=8$ million
pixels. A series of experiments using multiple reflections from a
curved mirror and a phase-shifting spatial light modulator array (SLM)
successfully demonstrated a variety of unitary transformations \citep{labroille_efcient_2014,morizur_programmable_2010}.
Fourier transforms can also be realized for other types of modes,
for instance waveguides modes and spectral temporal modes, in an efficient
manner \citep{lim_system_2011,lu_electro-optic_2018,lukens_frequency-encoded_2017}.
The FT method is the focus of this paper.

\begin{figure}
\includegraphics[width=1\linewidth]{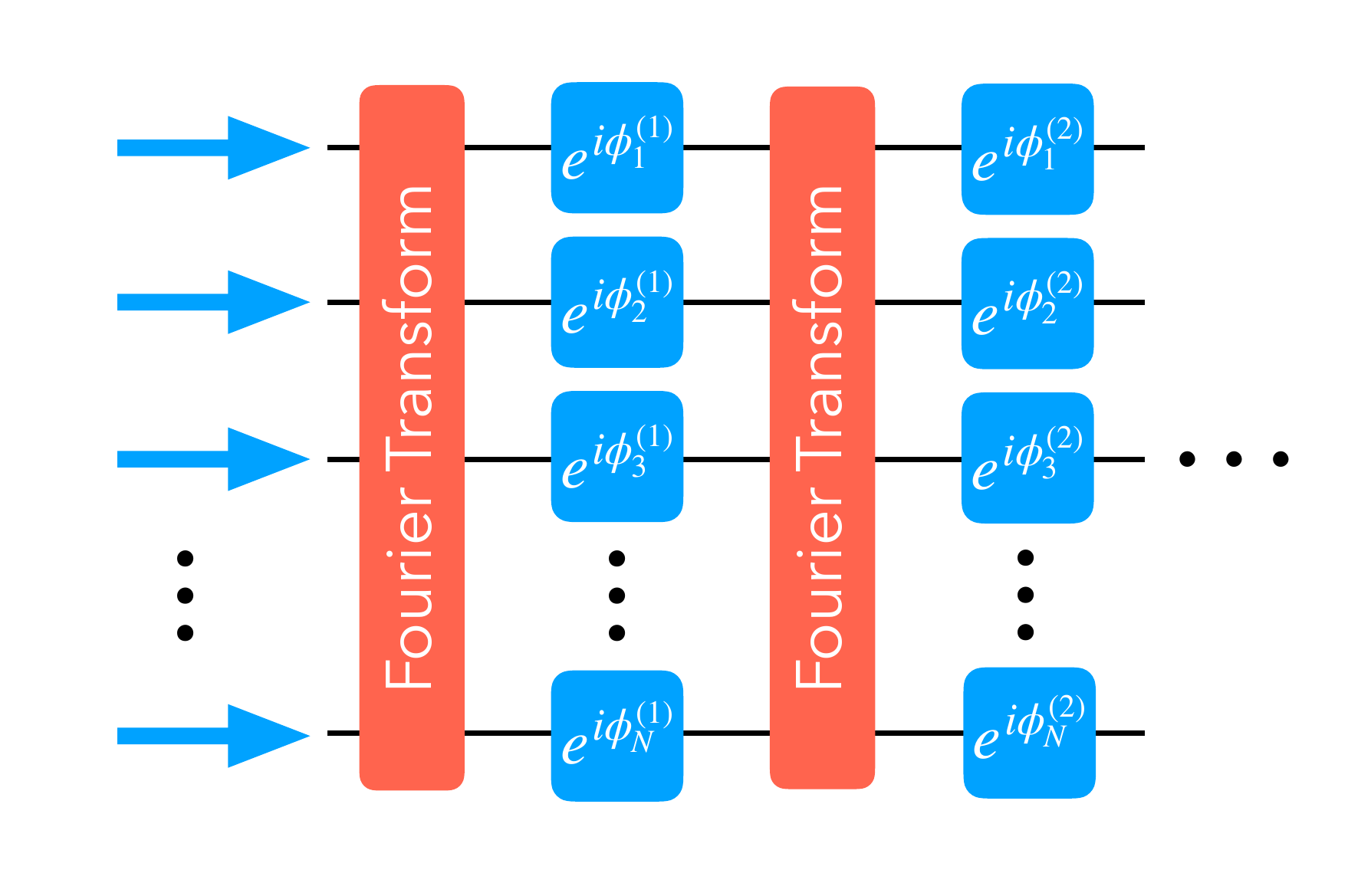}

\caption{In this work we show that any linear unitary transformation between
$N$ channels can be implemented by means of a succession of $6N+1$
phase masks (diagonal operators) and $6N$ Fourier transforms.}
\end{figure}

In particular, we give a deterministic algorithm to find the requisite
phase-shifts in the FT method. Rather than using the full continuous
FT, we use the discrete Fourier transform (DFT) since it is more amenable
to matrix algebra. While there is an existence proof showing that
a unitary could be decomposed into a sequence of FTs alternating with
phase-shifters \citep{borevich_subgroups_1981}, there is no prescription
for doing so with a sequence of realizable length. In \citep{schmid_decomposing_2000},
a method was found to decompose an arbitrary matrix as a sequence
of Fourier transforms and non-unitary diagonal matrices. However,
their method is not adequate for linear optical setups, since their
prescription makes use of non-unitary diagonal masks. Furthermore,
the length scales as $N^{3}$, which is very far from the optimal
scaling $L\sim N$. That said, an optimization algorithm to determine
these phase shifts, ‘wavefront matching’, was recently introduced
and experimentally validated \citep{fontaine_laguerre-gaussian_2019}.
While practical, iterative optimization has a number of drawbacks
for the FT method: 1. There is no guarantee of a solution nor its
global optimality 2. It does not prescribe the design parameters,
e.g., the required number of required FT-phase shift iterations, resolution
(i.e., SLM pixel size), and range (i.e., number of pixels). Thus,
it is unknown what is required to achieve a unitary of a given dimension,
level of optical loss, or amount of error. 3. Relative to the Reck
et al.’s deterministic algorithm, it is computationally slow. 4. It
does not provide physical insight into how to develop improved methods.
Consequently, there is a need for the deterministic algorithm we introduce
here.

In the first section, we map the Clements \emph{et al.} lattice onto
the FT method. That is, we decompose a layer of beamsplitters in the
lattice into a short sequence of FTs and fixed phase shifts. We use
this to adapt their deterministic algorithm to find the requisite
variable phase shifts in the FT method. Consequently, we give an explicit
prescription for how to design a unitary transformation with the FT
method. While the FT is often numerically computed using the DFT,
in the second section, we show that the DFT can also directly occur
in optical systems. We show that, for example, modal propagation in
a box waveguide is described by a DFT. In the appendix, we give a
detailed derivation of our FT method decomposition. More broadly,
the FT method is not restricted to optics, being directly applicable
to many other setups such as neutral atoms in optical traps or phonon
modes in ion chains.

\emph{Decomposition method.— }Any lossless, noiseless, linear transformation
on a closed system of $N$ optical modes is described by a unitary
matrix $U\in\mathcal{U}_{N}(\mathbb{C})$. Reck et al. showed that
any unitary transformation between optical modes can be implemented
as a lattice of beam splitters  \citep{reck_experimental_1994}. Such
a lattice is also known as a multiport interferometer. A beam splitter
is an optical element that mixes two modes $i$ and $j$  according
to unitary matrix $T(\theta,\phi)\in\mathcal{U}_{N}(\mathbb{C})$
parametrized by two angles $\phi,\theta\in[0,2\pi)$
\begin{equation}
\begin{pmatrix}\left[T(\theta,\phi)\right]_{ii} & \left[T(\theta,\phi)\right]_{ij}\\
\left[T(\theta,\phi)\right]_{ji} & \left[T(\theta,\phi)\right]_{jj}
\end{pmatrix}:=\begin{pmatrix}e^{i\phi}cos(\theta) & -sin(\theta)\\
e^{i\phi}sin(\theta) & cos(\theta)
\end{pmatrix}.\label{eq:bs}
\end{equation}
It  acts as the identity matrix on all the other channels. An arbitrary
beam splitter $T_{ij}(\theta,\phi)$ can be factorized in the following
way
\begin{equation}
\begin{pmatrix}e^{i\phi}cos(\theta) & -sin(\theta)\\
e^{i\phi}sin(\theta) & cos(\theta)
\end{pmatrix}=X\begin{pmatrix}e^{i\theta} & 0\\
0 & 1
\end{pmatrix}X\begin{pmatrix}e^{i\phi} & 0\\
0 & 1
\end{pmatrix},\label{eq:bs_decomp}
\end{equation}
where $X$ represents a 50-50 beam splitter, i.e., $X:=\frac{1}{\sqrt{2}}\begin{pmatrix}1 & 1\\
1 & -1
\end{pmatrix}$. Hence, one only needs controllable phase shifters and fixed 50-50
beam splitters to build the lattice of beam splitters designed by
Reck \emph{et al.}

Instead of a beamsplitter-based method, here we investigate a factorization
method based on Fourier transforms. As a starting point,we consider
the Discrete Fourier Transform (DFT), whose action is described by
a unitary matrix whose elements are given by $F_{jk}=\frac{1}{\sqrt{N}}e^{i2\pi jk/N}$.
Our design is built as a succession of Fourier transforms and phase
masks: 
\begin{equation}
U=D^{(0)}\prod_{i=1}^{L}FD^{(i)}.\label{eq:general}
\end{equation}
The phase masks $\left\{ D^{(i)}\right\} _{i\in\left\{ 0,..,L\right\} }$
are the only element in this setup that we can control. Each phase
mask on $N$ modes is described by a diagonal matrix parametrized
by $N$ angles, $D_{jk}^{(i)}=e^{i\alpha_{j}^{(i)}}\delta_{jk}$.
Thus, it is clear that one needs at least $N$ of them to simulate
an arbitrary multiport interferometer such as the Reck scheme.

\begin{figure}
\includegraphics[width=0.9\columnwidth]{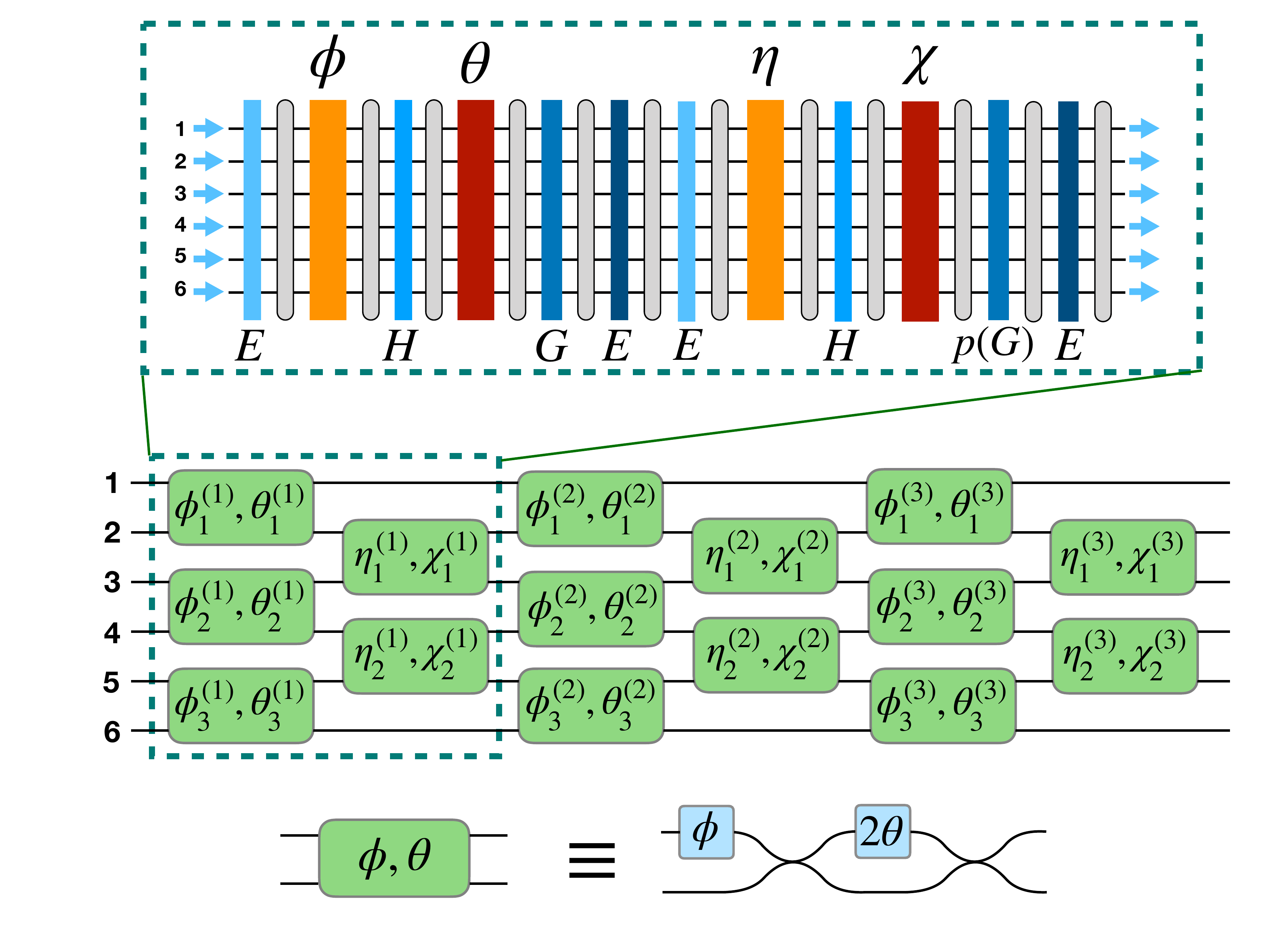}

\caption{\label{fig:diagram} Any unitary matrix can be realized by a mesh
of beam splitters, as described in \citep{clements_optimal_2016}.
Our decomposition method is based on replacing each layer of beam
splitters by a succession of discrete Fourier transforms and phase
masks, as schematically depicted here at the top. Each layer of beam
splitters requires six Fourier transforms (grey rounded rectangles)
and 6 phase-mask diagonal matrices (coloured rectangles). Only two
diagonal matrices per layer (red and yellow rectangles) depend on
the unitary matrix that is being implemented, while the rest (blue
rectangles) are fixed. The general expression for the phase masks
is given in Eqs(\ref{eq:b},\ref{eq:a},\ref{eq:cont}).}
\end{figure}

We present a way to find a decomposition of an arbitrary unitary matrix
in the form displayed in Eq.(\ref{eq:general}), consisting of $L+1=6N+1$
unitary diagonal matrices and $6N$ DFT matrices. In our factorization
method, rather than the Reck \emph{et al.} method we start from the
decomposition in beam splitters given by Clements \emph{et al.} in
\citep{clements_optimal_2016}. Their design consists of a mesh of
beam splitters arranged in $N$ consecutive layers. The composition
of the action of all the beam splitters in the mesh has the following
form 
\begin{align}
U & =D\prod_{i=1}^{N/2}\prod_{k=1}^{N/2-1}T_{2k}(\chi_{k}^{(i)},\eta_{k}^{(i)})\prod_{j=1}^{N/2}T_{2j-1}(\theta_{j}^{(i)},\phi_{j}^{(i)}),\label{eq:cl}
\end{align}
where $T_{j}(\theta,\phi)$ is a beam splitter mixing channels $j$
and $j+1$. As pointed out above in Eq.(\ref{eq:bs_decomp}), any
beam splitter can be implemented with two 50-50 beam splitters and
two phase shifters. Therefore, the design proposed in \citep{clements_optimal_2016}
can be understood as a succession of layers of 50-50 beam splitters
and phase masks. The procedure to translate the lattice of beam splitters
into a composition of phase masks and DFT's is schematically depicted
in Figure (\ref{fig:diagram}). In a nutshell, our decomposition builds
on this by factoring each layer of 50-50 beam splitters in the mesh
as a product of Fourier transforms and phase masks. Since the proof
is constructive, it automatically gives a method to find the parameters
in terms of $U$. Our decomposition needs only $N^{2}$ free controllable
parameters, which is optimal. However, an improvement by a constant
factor in the optical length (i.e. the total number of phase masks
required) may still be possible.

Consider that we want to decompose a given unitary matrix $U$. In
the decomposition displayed in Eq.(\ref{eq:cl}), each term of the
form $\prod_{k=1}^{N/2-1}T_{2k}(\chi_{k}^{(i)},\eta_{k}^{(i)})$ represents
a layer of beam splitters connecting each even channel $2j$ with
the odd channel $2j+1\Mod{N}$, whereas each term $\prod_{j=1}^{N/2}T_{2j-1}(\theta_{j}^{(i)},\phi_{j}^{(i)})$
represents a layer of beam splitters connecting each even channel
$2j$ with the odd channel $2j-1\Mod N$. It will turn out to be more
convenient for us if we relabel the indices as $\left\{ 0,1,2,3,...\right\} \rightarrow\left\{ 0,\frac{N}{2},1,\frac{N}{2}+1,...\right\} $.
Then, we can see that $U$ can also be decomposed as a succession
of layers of beam splitters such that in the odd layers each beam
splitter connects each channel $j\in\left\{ N/2,...,N-1\right\} $
with the channel $j-N/2$, whereas in the even layers each beam splitter
connects the channel $j\in\left\{ N/2,...,N-1\right\} $ with the
channel $j-N/2+1\Mod{\frac{N}{2}}$.

On the one hand, the odd layers can be written as $X\Xi{}_{1}^{(i)}X\Xi_{2}^{(i)}$,
where $\Xi_{1}^{(i)},\Xi_{2}^{(i)}$ are diagonal matrices. On the
other hand, the even layers have the same structure as the odd layers
after a cyclic shift of the first half of the channels. In other words,
each even layer of beam splitters can be expressed as $P^{T}X\Omega_{1}^{(i)}X\Omega_{2}^{(i)}P$,
where $\Omega_{1}^{(i)},\Omega_{2}^{(i)}$ are also diagonal matrices,
and $P$ is just the permutation matrix given by 
\begin{equation}
P_{jk}:=\begin{cases}
1 & k=j+1\,\mod\left(\frac{N}{2}\right),\,j\leq\frac{N}{2}-1\\
1 & k=j,\,j>\frac{N}{2}-1\\
0 & \textrm{{otherwise}}.
\end{cases}\label{eq:perm}
\end{equation}
Then, it follows that any unitary matrix admits the following decomposition
\[
U=D\prod_{i=1}^{N/2}\prod_{k=1}^{N/2-1}P^{T}X\Omega_{1}^{(i)}X\Omega_{2}^{(i)}P\prod_{j=1}^{N/2}X\Xi_{1}^{(i)}X\Xi_{2}^{(i)}.
\]

At this point, we only have to find how to decompose the matrices
X and $P$ as a product of phase masks and Fourier transforms. In
order to do this, we will find how to factorize them in products of
circulant and diagonal matrices. A circulant matrix is a matrix such
that each row is obtained by applying a cyclic shift by one slot to
the right to the previous row. Since any circulant matrix is diagonalized
by the DFT matrix $F$, a product of circulant and diagonal matrices
can always be re-expressed as a product involving only $F$, $F^{\dagger}$
and diagonal matrices.

Let us define the diagonal matrix $G:=\begin{pmatrix}I & 0\\
0 & iI
\end{pmatrix}$ and the circulant matrix $Y:=\frac{1}{\sqrt{2}}\begin{pmatrix}I & -iI\\
-iI & 1
\end{pmatrix}$. First, we note that $X=GYG$. Second, we observe that the permutation
matrix $P$ can be factorized as a product of three circulant matrices
and four diagonal matrices in the following way

\[
P=\frac{1}{\sqrt{2}}X\begin{pmatrix}C+I & C-I\\
C-I & C+I
\end{pmatrix}X,
\]
where $C$ is the cyclic shift matrix $C=\delta_{j,j+1\Mod{N/2}}$
of size $\frac{N}{2}\times\frac{N}{2}$. Therefore, we have shown
how to decompose any unitary matrix $U$ as a product of diagonal
and circulant matrices. If then we diagonalize the circulant matrices,
we immediately obtain a factorization of $U$ involving only $F$,
$F^{\dagger}$ and diagonal matrices. But the inverse of the DFT matrix
is just $F^{\dagger}=\Pi F=F\Pi$, where $\Pi$ is the following permutation
matrix
\[
\Pi_{jk}=\begin{cases}
1 & j=k=0\\
1 & j=N-k\\
0 & \textrm{otherwise}.
\end{cases}
\]
Since $\Pi D\Pi$ is diagonal whenever $D$ is diagonal, we can decompose
$U$ using only $F$ and diagonal matrices.

In the end, diagonalizing all the circulant matrices we obtain the
following expression

\[
U=DG\left[\prod_{i=1}^{N/2}B^{(i)}A^{(i)}\right]G{}^{\dagger},
\]
 where the terms $B^{(i)},A^{(i)}$ are given by

\begin{align}
B^{(i)} & =\left\{ E,p\left(G\right),H,\Gamma(\chi^{(i)}),E,p\left(G\Gamma(\eta^{(i)})\right)\right\} _{F},\label{eq:b}
\end{align}

\begin{align}
A^{(i)} & =\left\{ E,G,H,p\left(\Gamma(\theta^{(i)})\right),E,G\Gamma(\phi^{(i)})\right\} _{F},\label{eq:a}
\end{align}
where we made use of the notation 
\[
\left\{ D_{1},...,D_{N}\right\} _{F}:=\prod_{i=1}^{N}FD_{i}.
\]
The diagonal matrices $E,H$ are defined as

\begin{eqnarray*}
E_{jj}=\frac{1}{\sqrt{2}}\left[1-i\left(-1\right)^{j}\right],\\
H_{jj}=\frac{1}{2}\left[1-(-1)^{j}\right]+\frac{1}{2}\left[1+(-1)^{j}\right]e^{i2\pi j/N},
\end{eqnarray*}
and the diagonal matrix $\Gamma(\mathbf{v})$ is defined as a function
of a real vector $\mathbf{v}\in\mathbb{R}^{N/2}$:
\begin{equation}
\begin{array}{c}
\left[\Gamma(\mathbf{v})\right]_{jj}:=\begin{cases}
e^{iv_{j}} & j<\frac{N}{2}-1\\
i & j\geq\frac{N}{2}-1
\end{cases}.\end{array}\label{eq:cont}
\end{equation}
Finally, $p:\mathcal{U}_{N}\rightarrow\mathcal{U}_{N}$ is just the
map $p(U):=\Pi U\Pi$. Note that when applied on a diagonal matrix,
it just inverts the order of the diagonal entries after the first
one: $p\left(diag(a_{0},a_{1},...,a_{N-1})\right)=diag(a_{0},a_{N-1},...,a_{1}).$

We now summarize the procedure to create any unitary transformation
$U$ using phase-masks and DFTs. This procedure is based on using
Eqs.(\ref{eq:b},\ref{eq:a}) to express a unitary matrix according
to the factorization in Eq.(\ref{eq:general}). First, we permute
the channels of the unitary transformation as described in the proof,
which corresponds to computing the matrix $U_{P}=P^{T}UP$, with $P$
being the permutation matrix defined in Eq.(\ref{eq:perm}). Then,
we find the decomposition of $U_{P}$ as a lattice of beam splitters
by the procedure described in Clements \emph{et al.} \citep{clements_optimal_2016}.
That is, we find the parameters $\left\{ (\chi^{(i)},\eta^{(i)},\theta^{(i)},\phi^{(i)})\right\} _{i=\left\{ 0,...,N/2-1\right\} }$
for each lattice layer $i$ such that $U_{P}$ is factorized in the
form given by Eq.(\ref{eq:cl}). The procedure for finding these parameters
is explained in \citep{clements_optimal_2016}, but the general idea
is to null, one by one, all the off-diagonal elements of $U_{P}$
by means of an appropriate succession of beam splitters. We then apply
these parameters $\left\{ (\chi^{(i)},\eta^{(i)},\theta^{(i)},\phi^{(i)})\right\} _{i}$
as phase masks along with other fixed phase masks, all interleaved
with DFTs, to replace layer $i$. In Fig. \ref{fig:diagram}, we indicate
all seven different diagonal matrices (e.g., phase masks), $E,$ $H$,
$G$, $p(G)$, and $\Gamma(\mathbf{v})$ (labelled by the value of
$\mathrm{\mathbf{v}}=\chi,\eta,\theta,$ and $\phi$),  at the location
of their application within one layer $i$ of our method. All the
control parameters are contained in the diagonal matrices $\Gamma(\mathbf{v})$,
whereas the rest of the diagonal matrices are fixed. In the end, the
computation of all the diagonal matrices is quite efficient, as it
only requires $\mathcal{O}(N^{2})$ operations. In summary, applying
the structure in Fig. \ref{fig:diagram} in place of each the $N/2$
beamsplitter lattice layers results in an implementation of an arbitrary
unitary using only phase-masks and Fourier transforms.

\emph{Physical Implementation of the DFT.— }Our decomposition method
requires the capability to optically perform the DFT. Although the
standard continuous Fourier transform is routinely approximately performed
in optical experiments with several setups, such as lenses \citep{cutrona_optical_1960},
curved mirrors \citep{nikolov_fourier_1982}, or arrayed waveguide
gratings\citep{lim_system_2011}, implementing the DFT by optical
means is not a trivial problem. In waveguide systems a 'star coupler',
sometimes called a $N\times N$ symmetric multiport or splitter, is
sometimes said to perform a DFT-like operation in that any given input
mode is transformed to a flat distribution of output modes. That is,
the unitary matrix that describes the star coupler matches the magnitudes
of the DFT matrix, $|F_{jk}|$, but the phases will likely be incorrect.
Since setting the magnitude only removes half of the free parameters
in a unitary matrix, $N^{2}/2$ parameters still must be adjusted
to match a DFT. This cannot be accomplished by sandwiching the star
coupler between two phase-masks since they only have $2N$ parameters
together. Consequently, to implement our method there is a need for
an optical DFT in waveguide systems.

Here we discuss one possible procedure to optically compute the DFT
that is based on the phenomenon of self-imaging inside a multimode
waveguide \citep{Bachmann:94}. The idea of using multimode intereference
(MMI) couplers to realize the DFT is not new, and it was first proposed
in \citep{5471055}. However, their prescription uses an $2N\times2N$
MMI coupler to output two copies of the $N$ dimensional DFT on half
of the input modes, which is not amenable to our goal. Here we describe
a method to implement the DFT on $N$ modes with an $N\times N$ MMI
coupler, using all $N$ modes. As it only needs a planar waveguide
and phase shifts, we believe that our proposal could be easily scaled
to a large number of modes. Furthermore, it can be generalized to
many other setups, such as neutral atoms in optical traps.

\begin{figure}
\includegraphics[width=0.9\columnwidth]{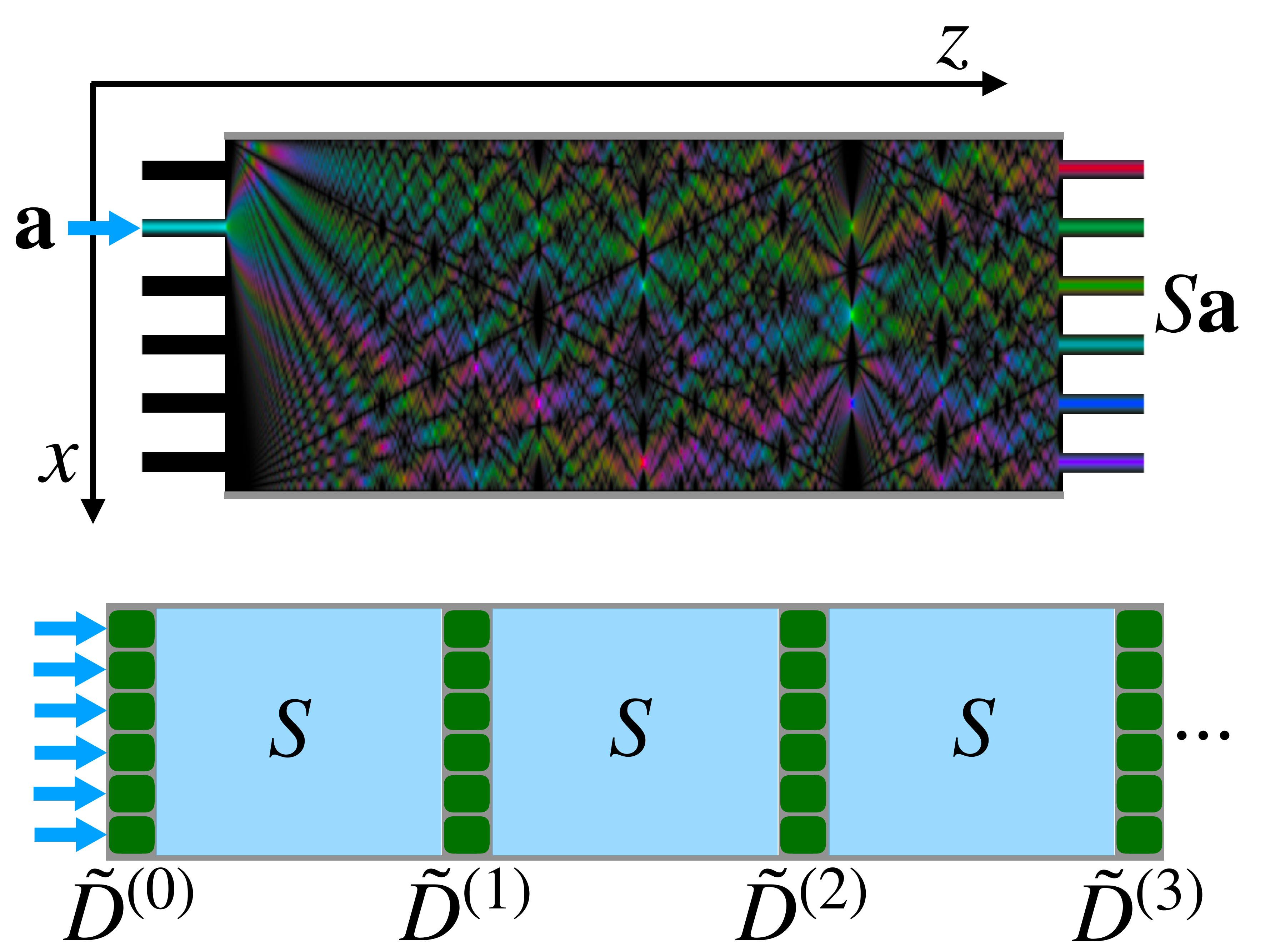}\caption{Free propagation inside a multimode waveguide can be used to realize
the DFT (in this case, a $6\times6$ DFT). When a wavepacket is injected
in the waveguide, the output at a particular value of the propagation
length is a superposition of $N$ copies of the wavepacket weighted
with complex phases. It can be shown that he output field is nothing
else than the DFT of the input field modulo phase masks and permutation
of the indices of the input and output modes. Therefore, a combination
of free propagation inside multimode waveguides and controllable phase
masks is enough to realize arbitrary unitary transformations. \label{fig:Free-propagation}}
\end{figure}

Consider a planar waveguide of width $w$ and index of refraction
$n$. We parametrize the transversal coordinate as $x$ and the longitudinal
coordinate as $z$. Let us assume hard wall boundary conditions, so
that it supports guided modes of the form $\psi_{n}(x)=\sin(k_{xn}x)$,
where $k_{xn}=\frac{\pi(n+1)}{w}$. Furthermore, let us assume that
the length of the waveguide is much larger than its width. Then, in
the paraxial limit we can approximate $k_{zj}=nk_{0}-k_{xj}^{2}/2nk_{0}$,
where $k_{0}:=\frac{2\pi}{\lambda}$.

Consider now that at $z=0$ we input a wavepacket $f(x-x_{j}^{in})$
centered at $x_{j}^{in}$. For simplicity, let us assume that $x_{j}:=\left(j+\frac{1}{2}\right)\frac{w}{N}$,
for some integer $j\in\left\{ 0,...,N-1\right\} $. This defines a
vector basis for our target DFT matrix in terms of $N$ wavepacket
modes. Under the assumptions listed above, it has been shown in \citep{Bachmann:94}
that when the propagation length is set to be equal to $z_{N}=\frac{2nk_{0}}{\pi N}w^{2}$,
the output field is given by 
\begin{equation}
E_{out}(x)=\frac{1}{\sqrt{N}}\sum_{k=0}^{N-1}e^{i\chi_{jk}}f\left(x-x_{k}^{out}\right).\label{eq:selfimaging}
\end{equation}
In other words, the output field is a superposition of $N$ repetitions
of the input wavepacket at $N$ distinct positions and weighted by
complex phases. The wavepackets $f\left(x-x_{k}^{out}\right)$ define
the output mode basis, where $x_{j}^{out}=\left(N-j-\frac{1}{2}\right)\frac{w}{N}$,
and the complex phase weights compose a unitary matrix, $S_{jk}:=\frac{1}{\sqrt{N}}e^{i\chi_{jk}}$.
In \citep{Bachmann:94} these weights were shown to be
\begin{equation}
S_{jk}=\begin{cases}
\frac{1}{\sqrt{N}}e^{i\frac{\pi}{4N}\left(k-j\right)\left(2N-k+j\right)+i\zeta_{0}} & \textrm{\textrm{if }}j+k\textrm{ is even}\\
\frac{1}{\sqrt{N}}e^{i\frac{\pi}{4N}\left(k+j+1\right)\left(2N-k-j-1\right)+i\zeta_{0}} & \textrm{\textrm{if }}j+k\textrm{ is odd,}
\end{cases}\label{eq:phases}
\end{equation}
where $\zeta_{0}:=-k_{0}z_{N}-\frac{\pi}{4}$. It is straightforward
to check that in fact the unitary matrix $S$ is nothing else than
the DFT matrix left and right multiplied by a diagonal matrix and
a permutation matrix 
\begin{equation}
S=R^{T}\Theta F\Theta R,\label{eq:dft_opt}
\end{equation}
where the permutation matrix $R$ and the diagonal matrix $\Theta$
are given by 
\[
R_{jk}:=\begin{cases}
1 & j\leq N/2\textrm{ and }2j-k-1=0\\
1 & j>N/2\textrm{ and }2j+k-2N-2=0\\
0 & \textrm{otherwise}
\end{cases},
\]
 
\[
\Theta_{jj}:=e^{i\frac{\pi}{4N}\left\lfloor \frac{j+1}{2}\right\rfloor ^{2}-i\frac{\zeta_{0}}{2}+i\pi\left\lfloor \frac{j+1}{2}\right\rfloor }.
\]

Eq.(\ref{eq:dft_opt}) implies a possible optical implementation of
the DFT. The setup would consist of a planar multimode waveguide with
length $z_{N}$ coupled to $N$ input channels and $N$ output channels
as in figure (\ref{fig:Free-propagation}). For an input field $E_{in}=\sum_{j=0}^{N-1}a_{j}f\left(x-x_{j}^{in}\right)$,
the output field is $E_{out}=\sum_{j=0}^{N-1}\tilde{a}_{j}f\left(x-x_{j}^{out}\right)$,
where the coefficients of the output are related to the coefficients
of the input by $\tilde{\mathbf{a}}=S\mathbf{a}$. 

Consider that we want to implement an arbitrary unitary matrix $U$
directly using such an MMI. Let us define the unitary matrix $U_{R}=RUR^{T}$.
We can use the previous results to find its factorization, $U_{R}=D^{(0)}\prod_{i=1}^{L}FD^{(i)}$.
Then, writing $F$ in terms of $S$ using Eq.(\ref{eq:dft_opt}),
the unitary matrix $U$ can be factorized as $U=\tilde{D}^{(0)}\prod_{i=1}^{L}S\tilde{D}^{(i)}$,
where we have defined the new diagonal matrices $\tilde{D}^{(0)}:=R^{T}D^{(0)}\Theta^{*}R$,
$\tilde{D}^{(i)}:=\Theta^{*}R^{T}D^{(i)}R\Theta^{*},\,i=1,...,L-1$,
and $\tilde{D}^{(L)}:=\Theta^{*}R^{T}D^{(L)}R$. Note that all matrices
$\tilde{D}^{(i)}$ are indeed diagonal matrices, since $R$ is just
a permutation matrix and $\Theta$ is also diagonal. In summary, to
use such an MMI in place of an exact DFT one simply needs to modify
the $L+1=6N+1$ phase-masks in our method.

This idea is not limited to optical modes in multimode waveguides.
In fact, any physical system with confined modes of the form $\psi_{n}(x)=\sin(k_{xn}x)$
and a parabolic dispersion relation $E_{j}=\frac{k_{j}^{2}}{2m}$
can be used to realize the DFT. In this case, instead of propagating
modes in a waveguide we consider a wavefunction that evolves inside
a rectangular well according to the Schrödinger equation. We start
with an input field of the form $\varphi(x,t=0)=\sum_{j=1}^{N}a_{j}\phi\left(x-x_{N-j+1}^{in}\right)$.
Now, the state at any time is given by $\varphi(x,t)=\sum_{n}c_{n}e^{iE_{j}t}\psi_{n}(x)$.
For free propagation, the dispersion relation is parabolic. Consequentially,
all the mathematical expressions are formally equivalent to the ones
that describe multimode interference in a waveguide. In particular,
one could apply this protocol to neutral atoms confined in an optical
trap.

Conclusion.— We have given an explicit, analytical, and deterministic
procedure to design an implementation of an arbitrary unitary transformation
of dimension $N$ using only discrete Fourier transforms and controllable
phase-masks. The number of control parameters is the minimum possible,
$N^{2}$. For mode sorting or multiplexing, where the global phase
of each output state is irrelevant, the required number of DFT-phase-mask
layers needed is $6N$. One additional phase-mask is required for
a completely arbitrary unitary. Thus, the scaling of the number of
layers with the dimension is linear and is optimal, up to an overall
factor. We have also prescribed the first practical method to implement
a DFT in integrated optics and even in systems outside optics, such
as ion traps. The unitary matrix factorization we give could also
be useful in quantum computation theory for decomposing quantum algorithms
in terms of just two types of operations, the quantum Fourier transform
and diagonal operators. We expect these results to be useful in a
variety of classical and quantum information applications in photonics
using various optical degrees of freedom including frequency-time,
orbital angular momentum, and position-momentum. 

\bibliographystyle{apsrev4-1}
\bibliography{OpticalUnitaries}

\end{document}